# Experimental examination of the alleged influence of cultural differences on single-file pedestrian fundamentals

*Nashiru Mumuni Daniel Bilintoh[1](belintonash@gmail.com), Jun Zhang[2*](junz@mail.ustc.edu.cn)*
State Key Laboratory of Fire Science, University of Science and Technology of China (USTC), Hefei 230027, China

**Abstract**

Multiple studies have addressed what influences the fundamental diagram in single-file experiments. Several studies indicate that there are differences by gender when group composition is taken into account. Even for the simplest systems, such as pedestrian flows in corridors, there are still some deficiencies in the detailed understanding of this fundamental relationship. The effect of cultural differences can only be seen in a different density and velocity interval if a comparison is conducted between the two different cultural groups. Our results indicate that the velocities of male and female pedestrians are different. For instance, when N = 20, the velocity is 0.65 to 0.69m/s for Chinese, while for Ghanaians, the velocity is 0.51 to 0.62m/s.

Similarly, when the densities change N = 40 in Chinese, they are 0.26 to 0.32m/s, while the Ghanaians are 0.21 to 0.28m/s. In addition, we investigated whether pedestrian behavior, corridor length, body weight, and height between Chinese and Ghanaian pedestrians influence cultural differences in a single-file fundamental diagram. Regression analysis makes it possible to determine whether there is a difference between the two groups.

# 1. Introduction

Analyzing pedestrian dynamics and their impact on human safety has been studied in several studies(Migon Favaretto et al., 2019)(Gulhare et al., 2018)(Banerjee et al., 2018). Understanding pedestrian dynamics is essential to maximize crowd control and plan facilities. In general, most crowds of people are composed of people who share a common interest and are gathered for cultural or recreational reasons. Despite a lack of interest in studies on cultural groups' impact on pedestrian dynamics, some have studied how individual difference affects pedestrian dynamics(Gulhare et al., 2018)(Subaih et al., 2020) (Biswal, 2014a).

The way fundamental diagrams (FDs) are originally conceived is to define the link between three factors: the density (1/m), the speed (m/s), and flow (Flötteröd & Lämmel, 2015)(Cao et al., 2017)(Cao et al., 2018). The FDs were initially developed to define the link between three factors. In (Zhang, n.d.), it is explained that FDs help describes pedestrian flows and densities and are related to numerous self-organizing occurrences in crowds. FDs have been discussed from a cultural perspective before. Divergent cultures and populations can also affect the flow, density, and speed of people (Thovuttikul et al., 2019)(Rodolfo Migon Favaretto et al., 2019)(Chattaraj et al., 2009b).

The present study examined Hofstede's dimensions of culture(Hall, 1966) (Hofstede, 2011). It proposed a method for analyzing videos according to the dimensions of Hofstede's theory of cognitive dimensions (Rodolfo M Favaretto et al., n.d.) and a method for analyzing crowd culture by applying the five-factor model to video collections. Use pedestrian tracking (Rodolfo M Favaretto et al., n.d.)(Rodolfo Migon Favaretto et al., 2017). The geometric representation of geometric features such as speed, angular variation, and distance allows them to be related to dimensions(Favaretto, R. M., Dihl, L., Musse, S. R., Vilanova, F., & Costa, 2017).

The illustrations of the speed-density relationship (or fundamental diagram) are enthusiastic about the cultural aspect. Different shapes have been obtained for the fundamental diagram by researchers doing experiments in different cultures(Rodolfo Migon Favaretto et al., 2019). People of different cultures walk differently, as this observation indicates. Across all pedestrian groups (Morrall, n.d.). Pedestrian speeds in Asian countries are slower than in western countries. According to(Tanaboriboon & Guyano, 1991), Singaporeans' walking speeds are slower than Americans,' but their flow rates are higher than those in western nations. (Tanaboriboon & Guyano, 1991) (Rahman et al., 2012) highlight the importance of reforming pedestrian design standards and integrating Western standards into this process. The pedestrian flow characteristics in Hong Kong are different depending on the type of walking facility, according to (Goh et al., 2012). According to (Rodolfo Migon Favaretto et al., 2019) (Chattaraj et al., 2009a), the behavior of Indians may be considered more like that of Middle Easterners. Having studied pedestrian movement across cultures (German and Indian), (Chattaraj et al., 2009a) concluded that density had less effect on the speed of Indian test subjects than German test subjects. Remarkably, Indians are more disorganized than Germans.

According to(Chattaraja et al., 2013), pedestrian flow in different cultures differs significantly from the fundamental diagram.

The primary purpose of this work is to investigate the impact of apparent social differences in pedestrian movement in a single-file fundamental between two countries, namely China and Ghana, using video recordings from experiments to automatically and manually extract the data. The studies have shown how pedestrian composition affects their movement characteristics based on density, speed, and path distance, considering gender, culture, and age. Compared the experiment results of mixed-gender groups in Ghana with similar conducted elsewhere, particularly in China, using single-file movement experiments(Cao et al., n.d.). Quantities of pedestrian movement, like speed changes and headway changes at different density levels, were used (Rodolfo Migon Favaretto et al., 2019) in India(Biswal, 2014a)(Gulhare et al., 2018), China(Ren et al., 2019)(Zhang et al., 2013), and Palestine(Subaih et al., 2020). Analyzing these data allows one to determine whether these parameters, such as flow, speed, and headway, differ between cultures.

## 2. Designed Experiment

An experiment involving a senior high school in the upper west region of Ghana was conducted in 2021. Figure 1 the scenario and the experiment are illustrated in the following image. The closed-loop corridor has a top circumference of 25.7m, two straight 5m sections, and two semicircles. (Ren et al., 2019)(Cao et al., n.d.)(Zhang et al., 2013). A measurement was made to ensure that overtaking would not occur during movement; the corridor was determined to be 0.8m wide. About 140 volunteers from Wa senior high school (ages between 17 and 24 years old) and 20 teachers from the school (ages between 30- and 65 years old) participated in the test. Findings from the study are illustrated in table 1. A total of 24 runs were conducted with the different types and numbers of participants in the corridor. All participants received the same instruction throughout each run. To collect more data, especially under steady-state, and ensure that the results are repeatable and meaningful, they were asked to walk normally for at least 3 minutes (free-flow) without overtaking. Taking into account the different numbers N of pedestrians in each run, the global densities $\rho_g = N/C$, 0.39 to 2.28m$^{-1}$ were the ranges for volunteers, and the same thing applied to the teachers and the mixed group because all the setup was the same number of pedestrians N = 10, 20,30, 40, 50 and 60.

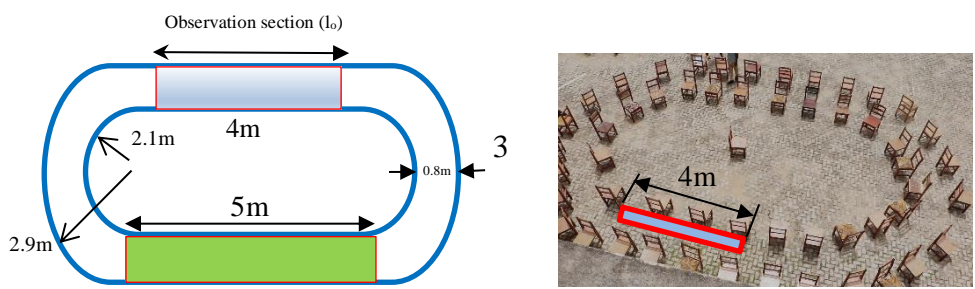

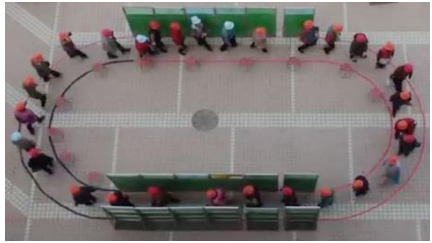 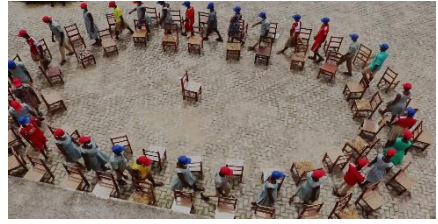

*(i) China set up reference (Ren et al., 2019)     (ii) Ghana setup*

*Fig. 1. Snapshots of the experiment (i) and (ii) representing Chinese and Ghanaians, respectively*

The experimental setup and measurement method for the Chinese study is identical to the one for the Ghanaian study; the only difference is the composition of the participants. During the experiment, we did not consider weather, temperature, and motivation as a factor. Manual and automatic extraction of the data (Biswal, 2014a) were used for both experiments. The adult ages were between 35 to 65, but that of the young ones was identical (Ren et al., 2019). The participants were 52 to 180cm, with an average height of 165cm.

| Ages | Males | Females | Total |
|---|---|---|---|
| 17 - 19 | 20 | 27 | 47 |
| 20 - 24 | 18 | 24 | 42 |
| 35 - 40 | 10 | 9 | 19 |
| 41 - 45 | 6 | 5 | 11 |
| 50 - 55 | 4 | 3 | 7 |
| 56 - 60 | 5 | 4 | 9 |
| 61 - 65 | 2 | 3 | 5 |
| Total | 65 | 75 | 140 |

*Table 1 Details of Chinese and Ghanaian participants and their composition*

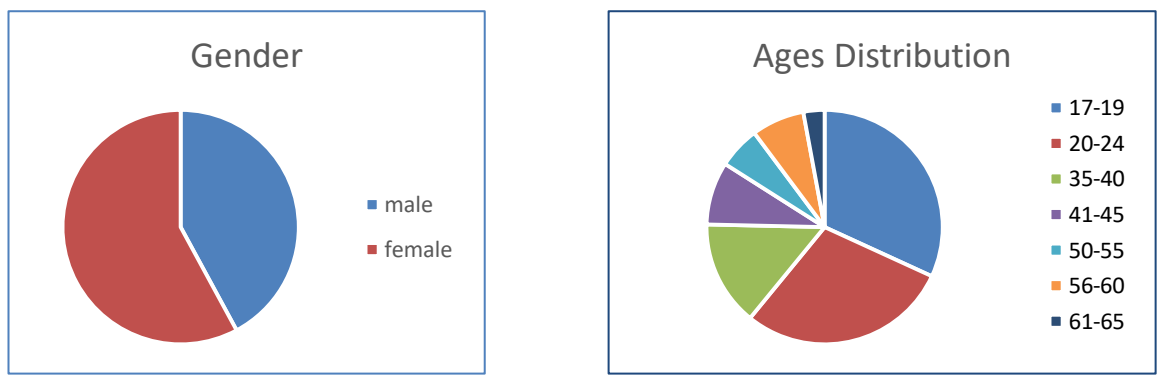

*Figure 2. Illustrate the gender and age distribution of Ghanaians in the experiments*

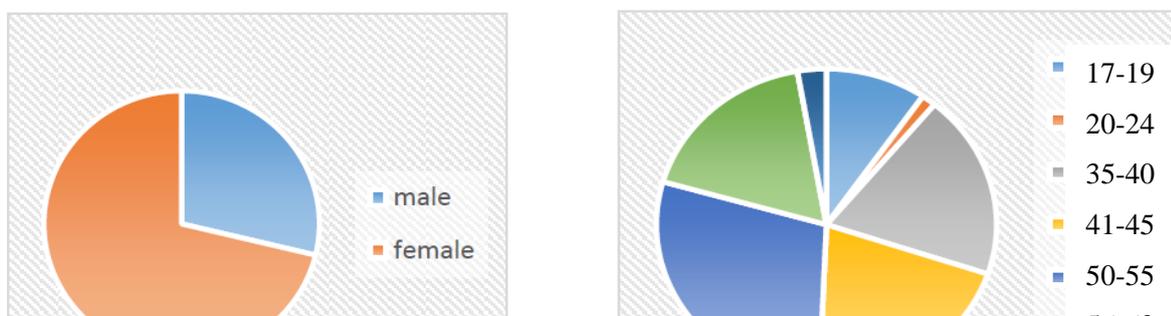

*Figure 3. The gender and age distribution of Chinese in the experiments was identical to* those (Ren et al., 2019). *only ages composition were different.*

## 3. Data

On the 3rd floor of a building, digital video cameras were mounted to collect pedestrian trajectories for the experiment. The recorded data was used to obtain information about time, speed, and density. We manually extracted pedestrian trajectories from the video recording (Biswal, 2014a). In the observation section ($l_o$), the velocity of individuals ($S_p$) is determined in the following way as referring (Biswal, 2014a):

$$S_p = \frac{l_0}{t_p^{out} - t_p^{in}} \quad (1)$$

Where $S_p$ is the velocity of the individual, $t_p^{out}$ is time out, $t_p^{in}$ is the time in, *and p* is the individual.

Now, average density, $k_p$ but we must find the density in a frame $K_f$. Which is also global density that is equation (1). $k_f = {N_f}/{L_o}$,

We know $N_f$ to be the pedestrian's number in the frame, *and* $l_o$ denotes the length of the observation section. Now to find $k_p$ from $k_f$ it will be:

$$K_p = \frac{\sum_{f=1}^{F} k_f}{F} \quad (2)$$

In the equation, $F$ is the total frame number (25) $k_f$ observed during the time interval $t_p^{out} - t_p^{in}$, and It can be attained as $\Phi \times (t_p^{out} - t_p^{in})$, with 25 frames/second as $\Phi$. the reciprocal of $K_p$ gives the average distance headway, referring to (30)(Biswal, 2014a).

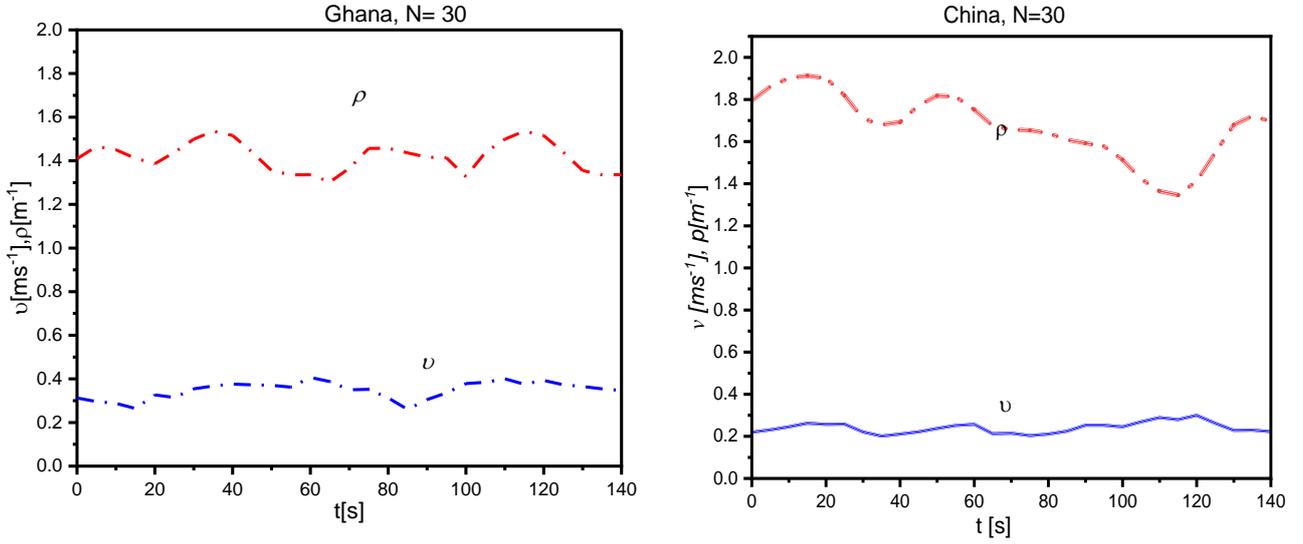

*Figure 4. Density and velocity development over time for the cycle with N = 30. Density is shown at the top of the graph; time slot and velocity v are shown at the bottom. Individual velocities fluctuate in correlation with density. After the passageway is opened at the end of the cycle, we see a change in the boundary conditions influenced by the starting phase. They are using the enhanced definition presented in Ref. (Seyfried et al., 2005) (Biswal, 2014a) for both the Ghanaian (left) and Chinese (right) cases.*

An illustration of how the density and velocity of individual subjects change over time can be found in figure 4. The pedestrians must adjust their initial movements because they do not react simultaneously. All the densities and velocities fluctuate slowly after the tuning phase. There is an apparent correlation between these 'microscopic' fluctuations. When the passageway opens, the density decreases, increasing velocity. This analysis excludes the data part where tuning phase influence and boundary condition rearrangements are evident.

### 3.1 Pre-processing of the data
Global density is used to calculate

$$\rho_{global} = N/L \quad (3)$$

In this case, N represents the pedestrian number. The number of pedestrians in our experiments was 10, 20, 30, 40, 50, and 60. The closed-loop corridor has a top circumference of 25.7 m with two straight sections of 5 meters and two semicircles each. Since the pedestrian number is 10 and the circumference is 25,7, global density equals pedestrian number (10) / circumference = 0.39 ped/m. By tracking pedestrians in manual modes, a pedestrian can be precisely determined by a standard video recording (25fps)(Seyfried et al., 2005). They did not reflect the actual position of the pedestrian; however, during movement, the body swings back and forth. Therefore, these trajectories are first smoothed based on step frequency (Liu et al., 2009) (pedestrian movements occur every 2 seconds). The smoothing window is one second, and the average is used.

| Index | Groups | Number of volunteers (male) | Number of volunteers (female) | Number teachers | Mixed group | Global density $\rho_g[m^{-1}]$ |
|---|---|---|---|---|---|---|
| 01 | Young | 10 | - | - | - | 0.39 |
| 02 | Young | 20 | - | - | - | 0.78 |
| 03 | Young | 30 | - | - | - | 1.17 |
| 04 | Young | 40 | - | - | - | 1.56 |
| 05 | Young | 50 | - | - | - | 1.94 |
| 06 | Young | 60 | - | - | - | 2.28 |
| 01 | Young | - | 10 | - | - | 0.39 |
| 02 | Young | - | 20 | - | - | 0.78 |
| 03 | Young | - | 30 | - | - | 1.17 |
| 04 | Young | - | 40 | - | - | 1.56 |
| 05 | Young | - | 50 | - | - | 1.94 |
| 06 | Young | - | 60 | - | - | 2.28 |
| 01 | Teach/v | - | - | 10 | - | 0.39 |
| 02 | Teach/v | - | - | 20 | - | 0.78 |
| 03 | Teach/v | - | - | 30 | - | 1.17 |
| 03 | Teach/v | - | - | 40 | - | 1.56 |
| 05 | Teach/v | - | - | 50 | - | 1.94 |
| 06 | Teach/v | - | - | 60 | - | 2.28 |

*Table 2 shows the details of experiment groups for both China and Ghana.*

## 4. Two culture comparisons of Time-Mean and Space-Mean Speeds

By taking the path that connects any two or more locations of interest, one can calculate their travel time immediately. The broad definition of travel time is "the amount of time required to travel between any two sites of interest." Travel time has two parts: running time, or the amount of time the means of transportation is moving, and stop delay time, or the amount of time it is still(Time & Collection, 1991). Figure 5 is a diagram that depicts the ideas of running time and paused delay time. Average or mean journey times are computed with formulae 3-6 and computer software. The average travel time estimates the total number of people that used the designated roadway during the period of interest.

According to Experiment 1, when comparing the time-mean and space-mean speeds between the two nations, China's transit time is 209.8 seconds while Ghana's is 226.6, showing that the Chinese fast more frequently than Ghanaians, who only fast around 33% of the time. Similar events occur in exp.6. However, the Chinese journey time is 7416.93 seconds longer than the Ghanaians' 6607.56 seconds. For experiment 1, the average Chinese time was 74.55 seconds, while Ghanaians' average time was 82.08 seconds. For experiment 6, the average Chinese time was 2448.93 seconds, while the Ghanaians' average time was 2287.84 seconds.

*i.     Table 3. Chinese Time-Mean and Space-Mean Speeds*

| Data | No. of ped. | Travel time (sec) | Running time (sec) | Delay time (sec) | Average travel speed (km/h) | Average running speed (km/h) | Sum | Average |
|---|---|---|---|---|---|---|---|---|
| Exp. 1 | 10 | 209.8 | 125.66 | 09 | 7.86 | 20.42 | 372.74 | 74.55 |
| Exp. 2 | 20 | 766.88 | 325.69 | 11 | 13.50 | 29.61 | 1146.68 | 229.34 |
| Exp. 3 | 30 | 1545.83 | 658.52 | 15 | 14.97 | 33.44 | 2267.76 | 453.55 |
| Exp. 4 | 40 | 3447.2 | 1370.77 | 20 | 11.42 | 29.79 | 4879.18 | 975.84 |
| Exp. 5 | 50 | 5583.58 | 2830.83 | 24 | 11.69 | 26.02 | 2892.54 | 1695.22 |
| Exp. 6 | 60 | 7416.93 | 4766.02 | 28 | 10.91 | 23.03 | 12244.9 | 2448.98 |

*The closed-loop corridor has a top circumference of 25.7 meters, 0.0257 km, and the observation section length is 4m (0.004 km). ped. (pedestrians)*

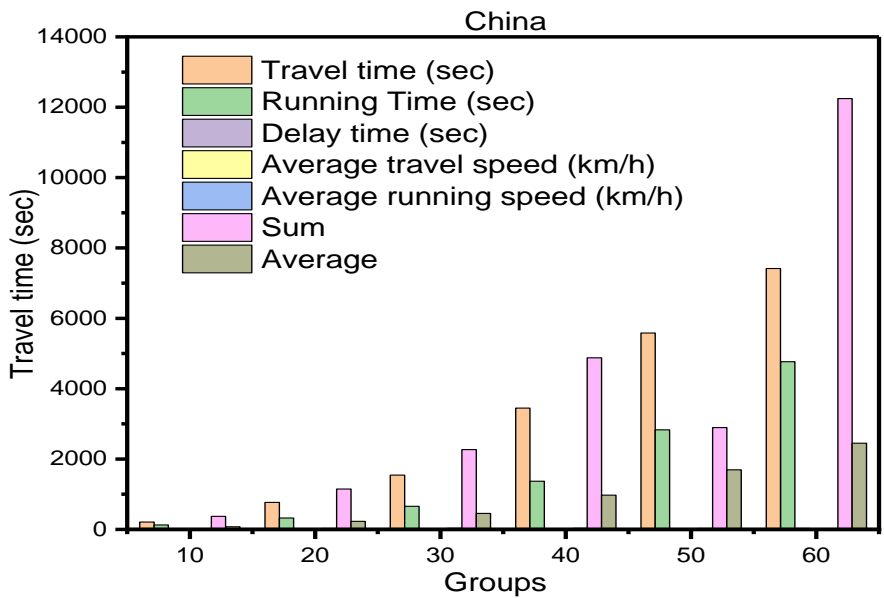

*Figure 5. An illustration of the time differences in Ghanaian pedestrians*

*ii.     Table 4. Ghanaians' Time-Mean and Space-Mean Speeds*

| data | No. of ped | Travel time (sec) | Running time (sec) | Delay time (sec) | Average travel speed (km/h) | Average running speed (km/h) | Sum | Average |
|---|---|---|---|---|---|---|---|---|
| Exp. 1 | 10 | 226.6 | 140.6 | 15 | 9.93 | 18.3 | 410.43 | 82.086 |
| Exp. 2 | 20 | 546.35 | 301.16 | 15 | 11.33 | 34.15 | 907.99 | 181.598 |
| Exp. 3 | 30 | 1051.81 | 689.08 | 20 | 10.72 | 33.57 | 1805.18 | 361.036 |
| Exp. 4 | 40 | 2211.76 | 1360.65 | 25 | 9.42 | 28.88 | 3635.71 | 727.142 |
| Exp. 5 | 50 | 4045.17 | 2830.83 | 30 | 8.67 | 22.82 | 6937.49 | 1387.498 |
| Exp. 6 | 60 | 6607.56 | 4766.02 | 33 | 10.07 | 22.53 | 11439.18 | 2287.836 |

*The closed-loop corridor has a top circumference of 25.7 meters, 0.0257 km, and the observation section length is 4m (0.004 km). ped. (pedestrians)*

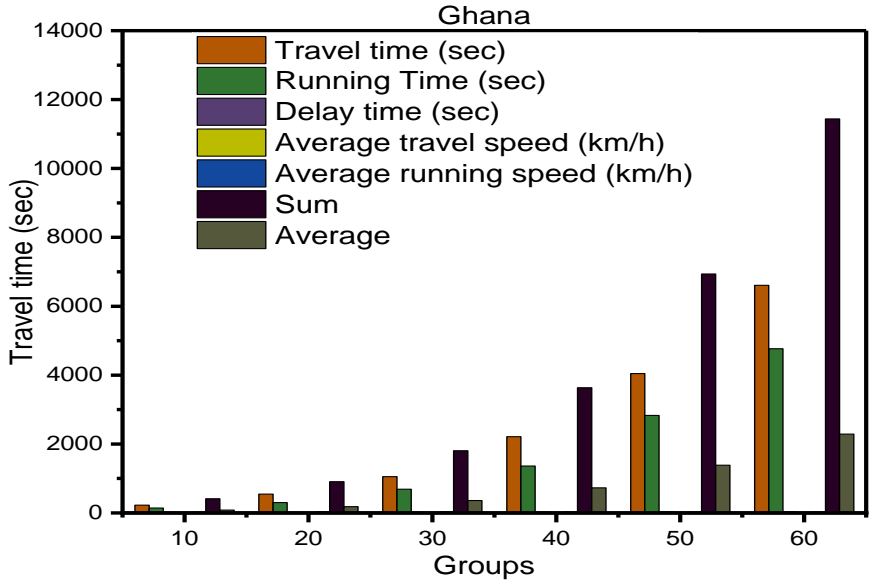

*Figure 6. An illustration of the time differences in Chinese pedestrians*

We are now calculating the difference between time-mean speed and space-mean speed. The estimated travel time (seconds)(Time & Collection, 1991)

$$\text{Estimated travel time (seconds)} = \frac{\text{Segment Lenght (km)}}{\text{Tune–Mean Speed } (\frac{km}{h})} \times 3{,}600 \text{ sec/hour} \quad (3)$$

$$Time - Mean\ Speed, \bar{v}_{TMS} = avg.\,speed = \frac{\sum v_i}{n} = \frac{\sum \frac{d}{t_i}}{n} \quad (4)$$

$$Time - Mean\ Speed, \bar{v}_{SMS} = \frac{distance\ traveled}{avg.travel\ time} = \frac{\frac{d}{\sum t_i}}{n} = \frac{n\ x\ d}{\sum t_i} \quad (5)$$

$$Average\ Running\ Speed, \bar{v}_r = \frac{distance\ traveled}{avg.running\ time} = \frac{d}{\sum \frac{t_{ri}}{n}} = \frac{n\ x\ d}{\sum t_{ri}} \quad (6)$$

*Therefore: $n$ is the number of observations, $d$ is the distance traveled or length, $i_t$ is the speed of the ith person, $i_t$ is the travel time of the ith person, and $t_{ri}$ is the running time of the ith person.*

Time-mean speed is the average of individual speeds, while space-mean speed (harmonic mean speed) is the distance traveled divided by the average travel time. Due to the longer time spent within the segment of interest, space-mean speeds weigh slow individual speeds more heavily. A relationship between time-mean speed and space-mean speed was developed by Wardrop (Wardrop, 1952) and illustrated by Equation 3-6 (Wardrop, 1952). According to the equation, only time-mean speed equals space-mean speed if the space-mean speed variance equals zero.

**5. The speed-density relation across gender**

The physiological characteristics of males and females differ in size, behavior, and reactions(Molnár, 2002). There was a mix of experiments (mixed group FMG) with female (FG) and male (MG) groups (Subaih et al., 2020). In terms of their movement characteristics in different compositions, this study aimed to determine if these differences affected them. A comparison of velocity-headway, velocity-density, and density-headway relationships was conducted. We calculated the fundamental quantities of parameters (density, velocity, and headway) based on the video recording using equation 1.

In figure 7. data were obtained from the mixed group (FMG), male group (MG), and female group (FG) experiments regarding density $\rho(t)$ and velocity $v(t)$.

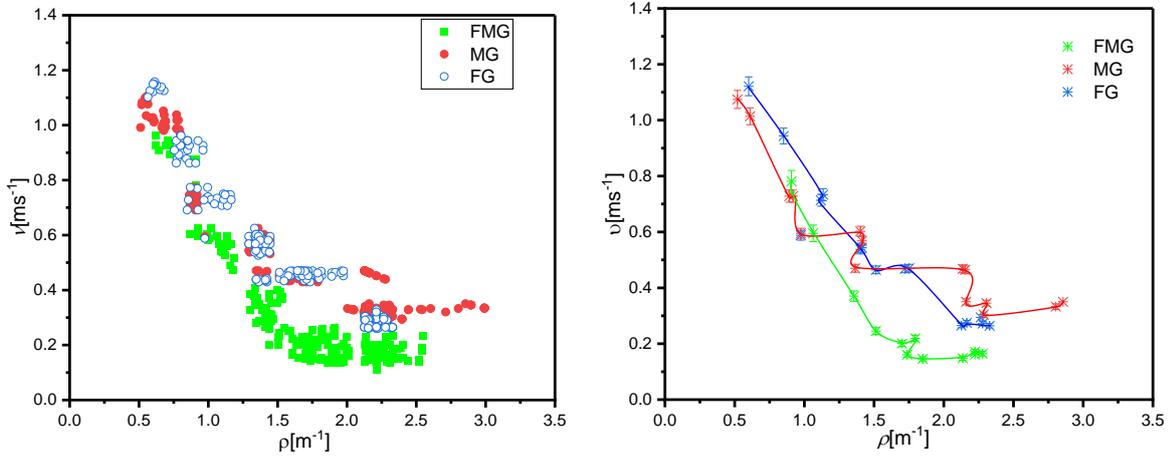

(i) All group's Density-velocity relation  (ii) All group's Density-velocity error bars

*Fig.7 illustrates the density-velocity relation for all groups and the error bars for different gender groups, mixed group (FMG), male group (MG), and female group (FG) for different densities and velocities.*

Parameters

|  |  | Value | Standard Error | t-Value | Prob>|t| |
|---|---|---|---|---|---|
| mixed | Intercept | 0.888 | 0.028 | 31.311 | 1.16117E-80 |
|  | Slope | -0.351 | 0.015 | -22.157 | 1.18944E-56 |
|  |  | Value | Standard Error | t-Value | Prob>|t| |
| female | Intercept | 1.219 | 0.016 | 74.208 | 1.25838E-151 |
|  | Slope | -0.437 | 0.010 | -43.357 | 3.92184E-106 |
|  |  | Value | Standard Error | t-Value | Prob>|t| |
| male | Intercept | 1.060 | 0.018 | 56.651 | 2.14165E-128 |
|  | Slope | -0.312 | 0.010 | -28.887 | 9.7856E-75 |

Summary

|  | Intercept |  | Slope |  | Statistics |
|---|---|---|---|---|---|
|  | Value | Standard Error | Value | Standard Error | Adj. R-Square |
| mixed | 0.888 | 0.028 | -0.351 | 0.015 | 0.700 |
|  | Intercept |  | Slope |  | Statistics |
|  | Value | Standard Error | Value | Standard Error | Adj. R-Square |
| female | 1.219 | 0.016 | -0.437 | 0.010 | 0.89 |
|  | Intercept |  | Slope |  | Statistics |
|  | Value | Standard Error | Value | Standard Error | Adj. R-Square |
| male | 1.06 | 0.018 | -0.312 | 0.010 | 0.799 |

*Table 5 shows density-velocity relation statistics of the mixed group (FMG), female group (FG), and male group (MG) from nonlinear fitting.*

Comparisons are based on FMG experiments with N = 10, 20, 30, 40,50, 60 pedestrians, MG with N = 10, 20, 30, 40,50, 60 pedestrians, and FG with N = 10, 20, 30, 40,50, 60 pedestrians. The essential diagrams are presented visually in Figure 7; we can observe the following details: density and speed show a negative correlation. Increased pedestrian volume in the corridor leads to a decrease in speed and walking speed. Second, the walking speed of men and women is similar in all density groups. Correspondingly, females' and males' velocities are 0.44, $\pm$ 0.1m/s, and 0.45$\pm$ 0.01m/s. However, when density is at 2.0m$^{-1}$, the average speed of the females and males walking is 0.23$\pm$ 0.1m/s and 0.25$\pm$ 0.07m/s, correspondingly. In table 5, the error values, slopes, and statistics $R^2$ are displayed for the mean velocities of the FG, the MG, and the FMG in different density ranges.

Furthermore, when mixed-gender groups walk together, pedestrians' velocity changes significantly if the density exceeds 0.50 m$^{-1}$. Within a mixed group of pedestrians walking at a median density of 1.56 m$^{-1}$, the average velocity is 0.45$\pm$0.1m/s. Given this, we conclude that walking groups are the same for men and women. Figure 8&9 shows that the velocities of males and females are the same at different densities. When walking in a mixed group, their velocities changes are insignificant (0.578$\pm$0.02m/s & 0.576$\pm$0.01m/s), as shown in figure 9. The figures show that both men and women reduced their walking velocity in mixed groups. Based on the mixed group's data analysis in figure 9, it appears that Ghanaian males and females prefer to generally walk in mixed groups and reduce their velocities than the Chinese.

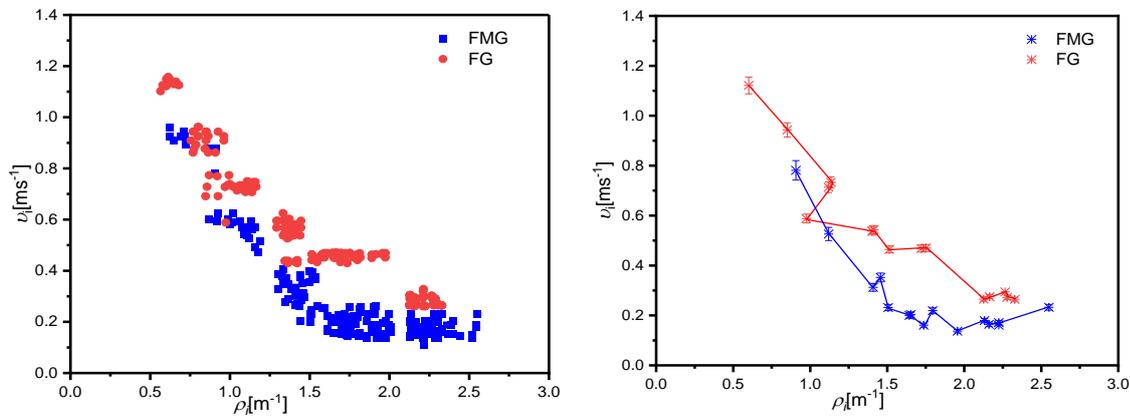

(i) Density-velocity on FMG and FG    (ii) Error bars on density-velocity

*Figure 8 Illustrating Density-velocity relation comparison between female group (FG) and Mixed group (FMG) experiments at different densities.*

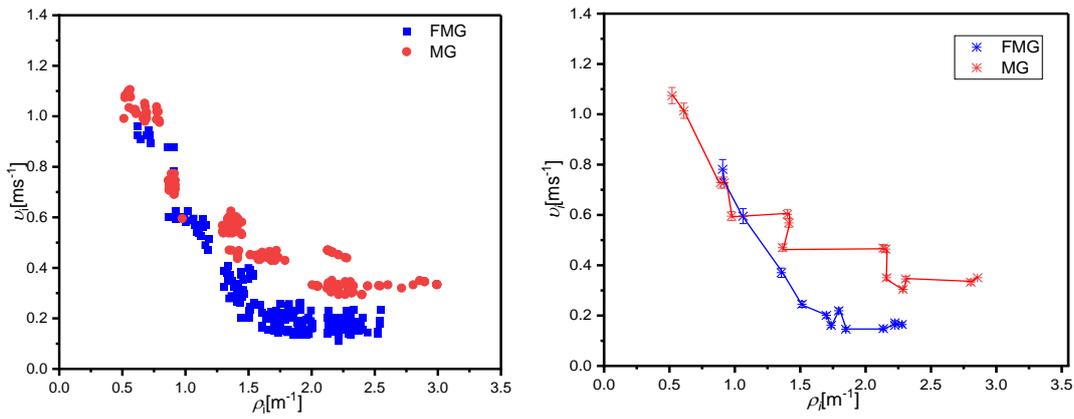

(i) Density-velocity on FMG and MG   (ii) Error bars on density-velocity

*Figure 9 Illustrating Density-velocity relation comparison between male group (MG and Mixed group (FMG) experiments at different densities.*

| | | Value | Standard Error | t-Value | Prob>|t| |
|---|---|---|---|---|---|
| velocity (m/s) | Intercept | 1.549 | 0.035 | 43.470 | 4.88777E-106 |
| | Slope | -0.413 | 0.016 | -24.400 | 8.04653E-63 |

| | Intercept | | Slope | | Statistics |
|---|---|---|---|---|---|
| | Value | Standard Error | Value | Standard Error | Adj. R-Square |
| velocity (m/s) | 1.549 | 0.035 | -0.413 | 0.016 | 0.740 |

*Table 6 shows density-velocity relation statistics from linear fitting.*

Pedestrians need a minimum amount of space based on their body depth and shoulder width. Fig 10. shows a single pedestrian standing on their feet with a simple rectangular body ellipse. Each pedestrian has a headway zone of 0.75m$^2$ (Biswal, 2014b). When walking, it is necessary to have space in front. The space in front of the path determines the speed of the trip, and it is also the space that determines the number of pedestrians that can pass a place in a given time.

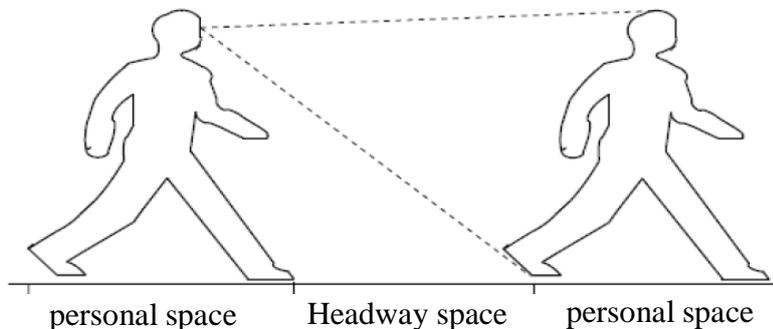

personal space   Headway space   personal space

*Figure 10 shows a sketch of a pedestrian keeping a specific distance from the other pedestrians(Biswal, 2014b).*

The headway transit system is the distance or duration between pedestrians. A minimum headway equals the shortest distance or time a system can achieve without reducing pedestrians' speed.

We also studied the 'headway' movement characteristic here. A sketch in figure 10 shows a pedestrian's specific distance from other pedestrians while walking to ensure they can move forward. This movement is analyzed in Figure 11 from the data obtained. If the pedestrian number rises in the corridor, then the headway of pedestrians drops for both genders. For example, a female has 1.3 $\pm$0.3m, and a male has 1.8$\pm$ 0.2m of headway for a density of 0.7m$^{-1}$. A reduction in space within the measurement section leads to different headway values with different densities, but the values are the same for different gender groups with similar densities.

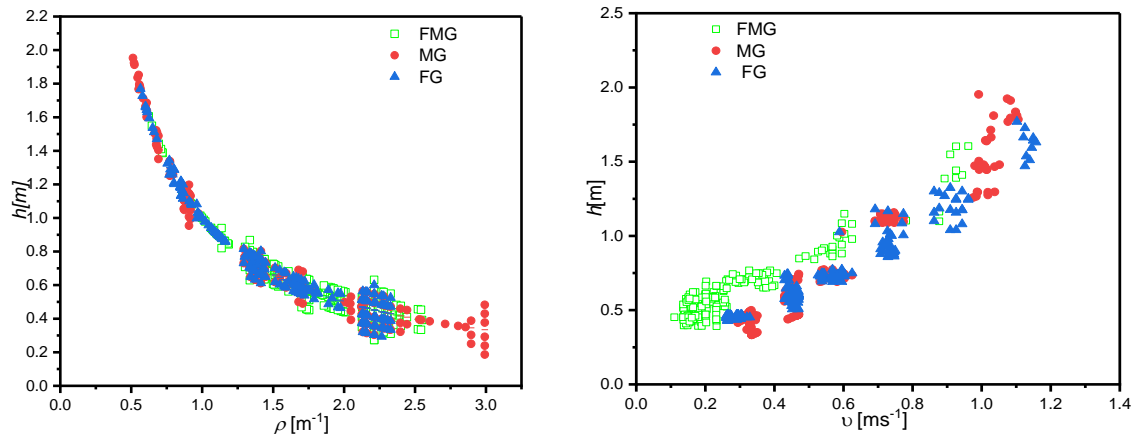

(a) Density-headway    (b) Velocity-headway

*Figures 11 (a) and (b) show headway density, velocity, and headway relation with different gender groups.*

## 6. Rate of overall flow between cultures

According to Fig. 12, flow rates were measured at 0.8 m throughout the experiment, with a total circumference length of 25.7m and 4 m (observation section). The pedestrians were divided by the time interval between the first and last pedestrian who passed the line to determine the flow rate. Slow-speed pedestrians' flow rate values were the least affected as they walked around the confined circular. Pedestrians' time and speed were observed from the observation section (4m) from the 25.7m circumference. Furthermore, as the number of pedestrians increases, slow speed occurs, but also their mixture causes impedance to average pedestrians, reducing their flow rate. Slow-speed pedestrians tend to exhibit this tendency more than fast-speed pedestrians.

In cases with increased pedestrians, the flow rate decreases less than in cases with fewer pedestrians. This way, average pedestrians must slow down because they can overtake slow-speed pedestrians. After all, the narrow space limits traffic on the path and reduces the density of the path.

Focus on the flow rate with the pedestrians in groups (10, 20, 30, 40, 50, 60) of the cases with a path width of 0.8m and observation section of 4m. Compared with a flow rate of 10 without slow-speed pedestrians, the Chinese pedestrian's flow rate was about 71% compared with Ghanaians' flow rate of 68% and 20 and 30. Both countries have similar percentages of 51% and a flow rate of 40 61% for Chinese and 60% for Ghanaians. The flow rate would be expected to be around 80% without slow-speed pedestrians; if the detour-distributed pedestrians caused the same effect as a single-lane block, the results in a more considerable experimental value. The results show that pedestrians walk in a predictable pattern, occupy a specific area of the path, and adjust their walking behavior according to circumstances, such as speed, distance, and walking course. At each density, the flow rate is either high or low, but with the increase to 60, the flow rate for both countries drops significantly due to the density difference and pedestrian culture factors. As a result, there is a high-density situation, and average pedestrians cannot overtake pedestrians who are strolling. However, the flow rate decrease in dense cases is slight with the increase in less dense cases. Consequently, average pedestrians cannot overtake slow-speed pedestrians since the path needs to be more comprehensive.

Experiments under the same conditions were repeated; however, the results were almost identical. The density, pedestrians' cultural beliefs, and the path's width strongly influence the flow rate.

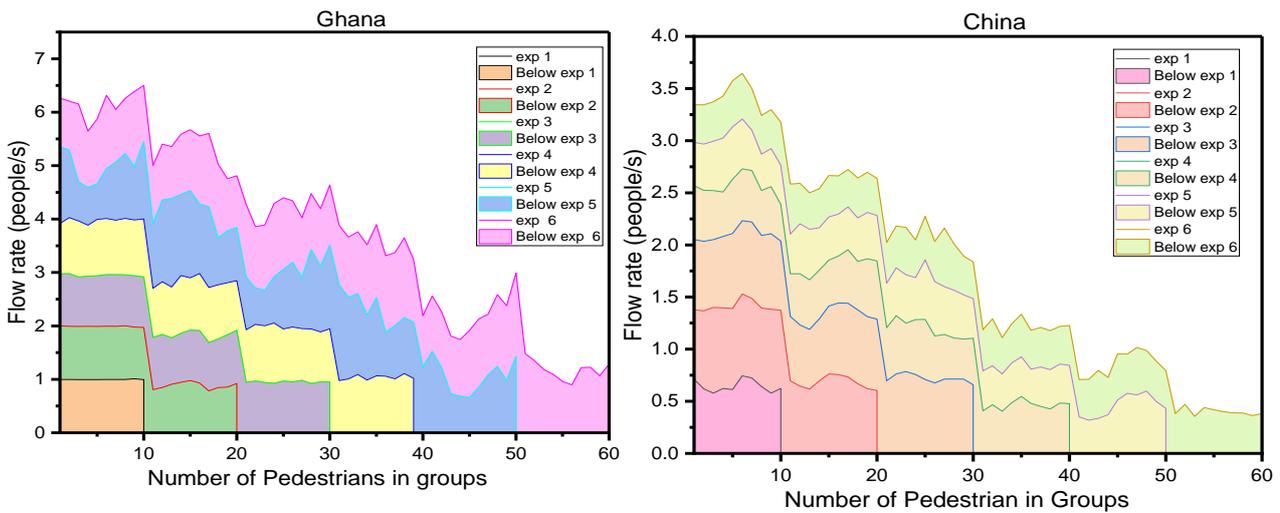

*Figures 12 (a) and (b) show the flow rate of pedestrians in different densities related to group numbers between the Chinese and Ghanaians.*

## 7. Speed-density relationship across cultures

Figure 13 below shows the velocity($v$)-density($\rho$) statistics for Chinese and Ghanaians under closed-corridor conditions. Below are some observations that emerge from looking at the data. First, the Chinese walk faster across several density zones than the Ghanaians, except at low density. Free-flow speed (low-density case) in Chinese $v_0^C$ = 1.21 ($\pm$0.11)m/s and 1.04 ($\pm$ 0.14)m/s for old and 1.07 ($\pm$ 0.15)m/s for the mixed group as shown in figure 9 and 10. is the same as the flow-free speed in Ghanaians $v_0^G$ = 1.16 ($\pm$0.12)m/s and 1.69 ($\pm$0.16)m/s, while Chinese data have been averaged over time, with more scatter than in Ghanaians, for example, N = 20, (N represent the number of participants) the range is 0.65 to 0.69m/s for Chinese while in Ghanaians the range is 0.51 to 0.62m/s. The density ranges for N = 40 in Chinese are 0.26 to 0.32$m^{-1}$, while in Ghanaians, they are 0.21 to 0.28$m^{-1}$. Hence, the differences. Although the Chinese composition of subjects is more homogeneous than that of Ghanaians, the ratio between individual passage times and distances to neighboring subjects is higher in the Chinese. According to the video recordings, this is the case. Chinese also appear to have a faster change in velocity with density changes (between 1 and 2$m^{-1}$) than the Ghanaians when densities are medium to high.

We also note that Chinese and Ghanaians have a linear relationship between density and velocity, as shown in Figure 14. In contrast, the density-velocity relationship (h) and velocity (v) shown in Figure 13 are more likely linear. To better understand this phenomenon, the h-v data for the two countries (China and Ghana) were obtained using linear fitting relationships and tested numerous hypotheses about the projected coefficients.

For Ghanaians and Chinese pedestrians, calculate individual mean densities and velocities. According to a mixed group (FMG) experiments, Ghanaians walk more slowly than Chinese pedestrians, as shown in Figure 9, with an average speed of 0.781m/s to 0.963m/s when there is low dense and 0.134m/s to 0.202m/s with high density, as compared to the Chineses of 0.99m/s to 1.035m/s and 0.187m/s to 0.233m/s respectively. The analysis revealed that Ghanaian pedestrians walk in low densities faster than Chinese pedestrians; the Chinese, however, walk in high densities in mixed groups faster than the Ghanaians. Based on our assumptions, these differences were caused by the alleged cultural influence and social conventions determining pedestrian comfort zones. According to this study, a comparison of groups of different ages from different cultures (Ren et al., 2019), people's walking movement is influenced by age and social diversity factors.

Compared to Ghanaian and Chinese pedestrians, when the velocity increases, they walk with more visible headway (Subaih et al., 2020)(Ren et al., 2019). Further, we can observe that the Ghanaians and Chinese walk at the same speed at a density equal to 1.3$m^{-1}$ and that when velocity is less than 0.7m/s, they walk at the same speed. Ghanaian pedestrians have a shorter headway than Chinese pedestrians when walking at speeds above 0.5m/s. Therefore, younger pedestrians walk with shorter headways, which vary depending on the alleged social diversity in the area.

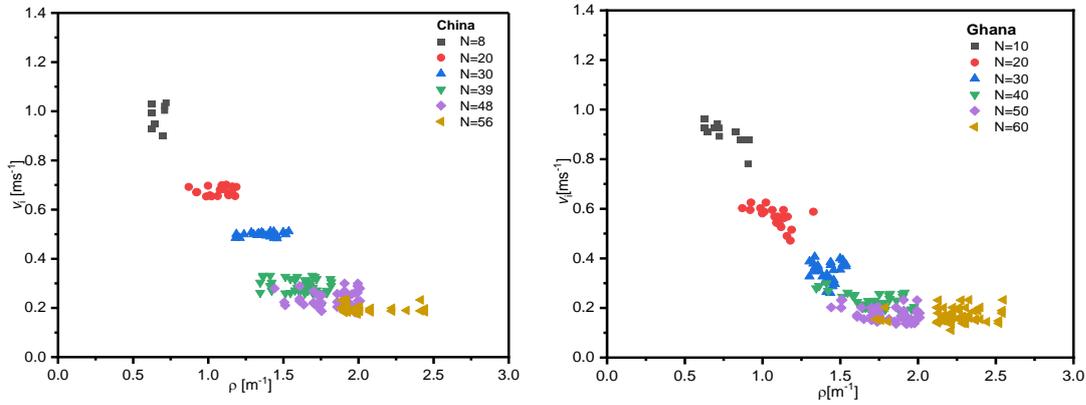

*Fig. 13. Illustrate Speed-density data for Chinese (left) and Ghanaian(right) diagrams separately.*

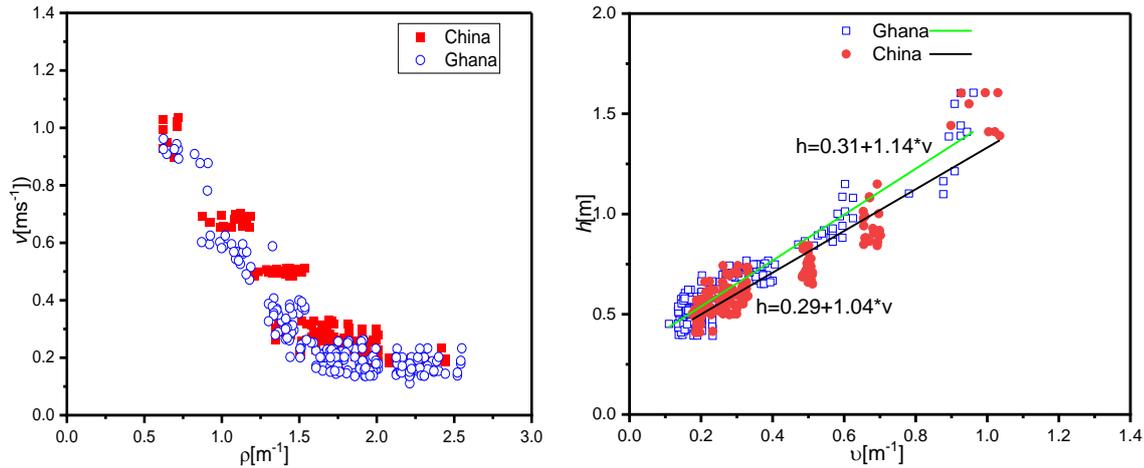

*Figure 14. Illustration of density-velocity distance relations and velocity-headway and their statistical analysis between Chinese and Ghanaians. Table 7 below details linear fitting and linear relation for headway-velocity $h = a + bv$ where h is the headway, $a$ is the intercept, $b$. is the slope, and $v$ is the velocity.*

## 8. Speed–distance headway relation across culture

An analysis of the Chinese and Ghanaian data is carried out using this equation $h = a + bv$, which is linearly fitted for the analysis (Biswal, 2014a). Table 7 shows the values of $a$ and $b$ for the Chinese and Ghanaian data, and Statistics are also provided. Linear fitted data are also displayed, along with other statistics shown in figure 13. A statistically linear fit is obtained using t-statistics and $R^2$ values in tables 7 (i) and (ii). However, $a$ represents the personal space (intercept), whereas $b$ represents the sensitivity in the relationship(slope) $h = a + bv$, h to $v$, or $\frac{dh}{dv}$ sensitivity.

(i) Parameters for china

|  |  | Value | Standard Error | t-Value | Prob>|t| |
|---|---|---|---|---|---|
| China | Intercept | 0.291 | 0.011 | 26.035 | 2.59185E-66 |
|  | Slope | 1.040 | 0.027 | 37.243 | 3.65732E-92 |
| Ghana | Intercept | 0.309 | 0.008 | 35.412 | 9.72952E-91 |
|  | Slope | 1.144 | 0.025 | 44.503 | 3.34128E-109 |

(ii) Summary

|  | Intercept | Slope | Statistics |
|---|---|---|---|

|        | Value | Standard Error | Value | Standard Error | Adj. R-Square |
|--------|-------|----------------|-------|----------------|---------------|
| China  | 0.291 | 0.011          | 1.040 | 0.027          | 0.872         |
| Ghana  | 0.309 | 0.008          | 1.144 | 0.025          | 0.903         |

(iii) ANOVA

|       |       | DF  | Sum of Squares | Mean Square | F Value   | Prob>F      |
|-------|-------|-----|----------------|-------------|-----------|-------------|
|       | Model | 1   | 8.903          | 8.903       | 1387.052  | 3.65736E-92 |
| China | Error | 201 | 1.290          | 0.006       |           |             |
|       | Total | 202 | 10.193         |             |           |             |
|       |       | DF  | Sum of Squares | Mean Square | F Value   | Prob>F      |
|       | Model | 1   | 10.068         | 10.068      | 1980.533  | 3.34135E-109|
| Ghana | Error | 211 | 1.072          | 0.005       |           |             |
|       | Total | 212 | 11.141         |             |           |             |

*Table 7(i). (ii), (iii) show China's velocity-headway parameters, summary, and ANOVA*

Additionally, We examined 'headway' movement characteristics in both countries—an example of how a pedestrian maintains a specific distance while walking is shown in figure 10. The data obtained are used to analyze this movement in figure 14, and a details explanation is in table 7. There is a reduction in pedestrian headway when both countries' numbers rise in the corridor. However, the headway of a Chinese is 1.04±0.2m, and that of Ghanaians is 1.14±0.2m, giving that density at the medium stage. Despite differences in headway values between the two countries with similar densities because of reduced space within the measurement section, the Ghanaians' values are higher than the Chinese, which may be attributed to cultural factors, with statistics $R^2$ of 0.87 and 0.90, respectively.

## 9. Effect of Personal Space/ Body Size of Pedestrians movement

We compare body size and personal space to determine if they influence the pedestrian's trajectories. Based on this information, we calculated both countries' average body mass index (BMI). For Chinese, the BMI is 23.10kg/m$^2$ and 26.34kg/m$^2$ for young students and older adults (No et al., 2021)(Maitiniyazi et al., 2021), respectively, and personal space is about 0.29m. We use the mean of the former two sizes, 36m, for the mixed group with a standard error of 0.01. Whiles the average body mass index (*BMI*) for Ghanaians is between 25kg/m$^2$ to 29 kg/m$^2$ for males and females(Tuoyire et al., 2018)(Neupane et al., 2016) with standard errors of 0.02, and personal space is about 0.36m. but the average personal space per person is about 0.7m$^2$(Biswal, 2014a). This hypothesis claims that China, $a^C$, and Ghana, $a^G$, have equal minimum personal space (Null hypothesis $H_0$: $a^C - a^G = 0$ and alternate hypothesis $H_a$: $a^C - a^G \neq 0$). we conducted testing. We consider many data points from experiments, assuming the expression would be standard normality.

|       |              | Value | Standard Error | t-Value | Prob>|t|     |
|-------|--------------|-------|----------------|---------|-------------|
| China | a(m) Intercept | 0.291 | 0.011        | 26.035  | 2.59185E-66 |
| | b(s). Slope | 1.040 | 0.027 | 37.243 | 3.65732E-92 |
| Ghana | a(m). Intercept | 0.358 | 0.016 | 21.916 | 2.16755E-56 |
| | b(s). Slope | 0.945 | 0.046 | 20.201 | 2.61559E-51 |

*Table 8 statistical measures for headway-velocity (h-v) relationship*

$$z = \frac{a^C - a^G}{\sqrt{S_{a^C}^2 + S_{a^G}^2}} \tag{7}$$

where $z$ is the critical value, $a^C$, and $a^G$ represent Chinese and Ghanaian, respectively, and $S_{a^C}$ and $S_{a^G}$ are the standard error for the data from Chinese and Ghanaians in table 6 above.

A normal distribution of the data is reasonable to assume. For the Chinese, $S_{a^C}$ is the standard error, and for Ghanaians, $S_{a^G}$ is the standard error. May reject the null hypothesis by calculating the $z$ value as more significant than the critical value. Otherwise, it cannot. According to the expression above, the $z$ critical value for a two-tailed test of the null hypothesis is 2.47. Because the z-critical two-tailed test values are more significant than 1.96, the null hypothesis is not valid. Hence, the estimated minimum personal space in Ghanaians and Chinese differ statistically. In Ghanaian, it is more significant than in Chinese. As a result, the jam density in Chinese seems higher than that of Ghanaians. This study is not affected by body mass index in any significant way.

## 10. Effect on the corridor length

This graph displays headway distance (h) and speed (v) in Ghana's closed corridors of different distances $l_p$ = 25.7m and 34.5m in Figure 15. The double-length corridor needs to be constructed at comparable densities; we selected N = 10, 20, 30, 40, 50, and 60 test subjects for the fitted relation $h = a + bv$; table 9 shows values for $a$, $b$, and $R^2$. Hypotheses tests comparing intercepts and slope terms between the short and long-length sections did not show significant differences. Therefore, the distance of the corridor has no significant influence on headway-velocity relations.

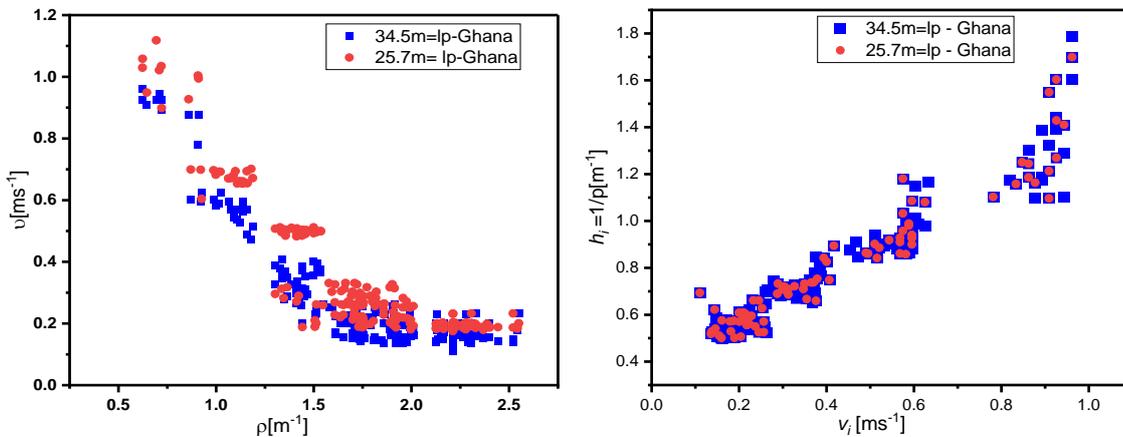

*Figure 15. Illustrates different lengths of corridor experiment data for density-speed and speed-headway performed in Ghana*

(i) parameters

(i)

|  |  | Value | Standard Error | t-Value | Prob>|t| |
|---|---|---|---|---|---|
| $l_p$ = 25.7m | a[m]-Intercept | 0.357 | 0.012 | 27.613 | 2.42786E-64 |
|  | b[s]-Slope | 1.061 | 0.025 | 41.118 | 1.32253E-89 |
| $l_p$ = 34.5m | a[m]-Intercept | 0.367 | 0.022 | 27.733 | 2.51386E-64 |
|  | b[s]-Slope | 1.074 | 0.034 | 41.434 | 1.41153E-89 |

(ii) summary

|  | Intercept | | Slope | | Statistics |
|---|---|---|---|---|---|
|  | Value | Standard Error | Value | Standard Error | Adj. R-Square |
| $l_p$ = 25.7m | 0.357 | 0.012 | 1.061 | 0.025 | 0.909 |
| $l_p$ = 34.5m | 0.367 | 0.022 | 1.062 | 0.034 | 0.919 |

(iii) ANOVA

|  |  | DF | Sum of Squares | Mean Square | F Value | Prob>F |
|---|---|---|---|---|---|---|
| $l_p$ = 25.7m $l_p$ = 34.5m | Model | 2 | 12.089 | 12.089 | 1690.696 | 1.32256E-89 |
|  | Error | 168 | 1.201 | 0.007 |  |  |
|  | Total | 169 | 13.290 |  |  |  |

*Table 9 shows the statistical analyses of the relationship in different corridor lengths.*

## 11. Conclusion

Specifically, a comparison of fundamental diagrams from two cultures is made in this study, namely Chinese and Ghanaians. We present two different lengths of corridors as well as their fundamental diagrams. Additionally, we considered body size and minimum space estimates. All measurements used the same method and statistics, such as t-tests and z-tests, to compare results. We have noticed the following: There was a difference in the free flow speed and time between the Chinese and Ghanaian groups, suggesting similar behavior when moving independently. Compared to the Chinese group, the headway increase with acceleration is more significant for the Ghanaians group, indicating that the Chinese are less sensitive to density increases than the Ghanaians. A cultural difference may be one reason for these differences.

We also examined the composition and dynamics of pedestrian movement. Studying how Chinese and Ghanaians move in single-file based on different pedestrian characteristics. For instance, when N = 20 (N represents the number of participants), the velocity is 0.65 to 0.69m/s for Chinese, while for Ghanaians, the velocity is 0.51 to 0.62m/s. The density ranges for N = 40 in Chinese are 0.26 to 0.32$m^{-1}$, while in Ghanaians, they are 0.21 to 0.28$m^{-1}$. Our study found that gender does not affect pedestrian movement when both men and women are present separately, but when mixed groups are included, the velocity changes. When mixed-gender pedestrians walked together, their velocities began to slow because of social conventions.

We also observed that the Chinese have a faster flow rate across several density zones than the Ghanaians, except at low density. Compared with a flow rate of 10 without slow-speed pedestrians, the Chinese pedestrian's flow rate was about 71% compared with Ghanaians' flow rate of 68% and flow rate of 20 and 30. Both countries have a similar percentage of 51% and a flow rate of 40, 61% for Chinese, and 60% for Ghanaians. According to the data analysis, this is the case. Chinese also appear to have a faster change in velocity with density changes (between 1 and 2m-1) than the Ghanaians when densities are medium to high.

As well, this paper could not analyze the age factor. However, we observed that the elderly individuals walked slower when densities differed. Comparable Chinese pedestrians were faster than Ghanaian pedestrians in densely populated areas. There are several interesting and important facts that these results have revealed about how gender influences pedestrian movement: it is generally believed that female pedestrians move differently than male pedestrians because of sociocultural influences. Our study cannot confirm this assumption, as female and male participants exhibit the same features. More experimental data is needed for work on lower and higher densities for both males and females in different age compositions and countries, especially in Africa, where we have different cultural beliefs.

**Acknowledgments Funding**
This publication acknowledges all sources of funding for the work described.:
The authors acknowledge the foundation support from the National Natural Science Foundation of China (Grant No. U1933105, 72174189) and the Fundamental Research Funds for the Central Universities (Grant No. WK2320000043, WK2320000050).